\documentclass[%
 reprint,
 amsmath,amssymb,
 aps,
]{revtex4-2}

\usepackage{steinmetz}
\usepackage{graphicx}
\usepackage{dcolumn}
\usepackage{bm}

\usepackage{changes}
\usepackage{amsmath}
\usepackage{comment}
\usepackage{color,soul}
\usepackage{hyperref}       
\hypersetup{
    colorlinks=true,
    linkcolor=blue,
    filecolor=magenta,      
    urlcolor=cyan,
}
\begin{document}

\preprint{APS/123-QED}
\title{Voltage-tunable OPO with an alternating dispersion dimer integrated on chip}


\author{Dmitry Pidgayko$^{1,2}$}
\author{Aleksandr Tusnin$^{1,2}$}
\author{Johann Riemensberger$^{1,2}$}
\author{Anton Stroganov$^{3}$}
\author{Alexey Tikan$^{1,2}$}%
 \email{alexey.tikan@epfl.ch}
\author{Tobias J. Kippenberg$^{1,2}$}%
 \email{tobias.kippenberg@epfl.ch}
\affiliation{%
 $^1$Institute of Physics, Swiss Federal Institute of Technology Lausanne (EPFL), Lausanne, Switzerland\\
 $^2$Center for Quantum Science and Engineering, EPFL, CH-1015 Lausanne, Switzerland\\
 $^3$LIGENTEC SA, EPFL Innovation Park, CH-1024 Ecublens, Switzerland 
}%

\date{\today}

\begin{abstract}
Optical parametric oscillators enable the conversion of pump light to new frequency bands using nonlinear optical processes. Recent advances in integrated nonlinear photonics have led to the creation of compact, chip-scale sources via Kerr nonlinearity-induced parametric oscillations. While these sources have provided broadband wavelength tuning, the ability to tune the emission wavelength via dynamically altering the dispersion, has not been attained so far.  Here we present a voltage-tunable, on-chip integrated optical parametric oscillator based on an alternating-dispersion dimer, allowing us to tune the emission over nearly 20 THz near 1550 nm. Unlike previous approaches, our device eliminates the need for a widely tunable pump laser source and provides efficient pump filtering at the drop port of the auxiliary ring. Integration of this scheme on a chip opens up the possibility of compact and low-cost voltage-tunable parametric oscillators with diverse application possibilities.
\end{abstract}

\maketitle

\section{Introduction}
Nonlinear optical processes are of paramount importance in both science and technology, ranging from supercontinuum generation that unlocked frequency metrology to squeezed light generation as used in gravitational wave detection or optical parametric oscillators that generate mid IR radiation for spectroscopy~\cite{kippenberg2018dissipative, kippenberg2011microresonator}.
Advances in integrated nonlinear devices have enabled to access traditionally weak nonlinear optical processes with continuous wave lasers, at only milli-Watt of pump power levels and triggered the development of chip-scale optical frequency combs, i.e. microcombs~~\cite{chang2022IntegratedOpticalFrequency,hermans2022OnchipOpticalComb}. These devices utilize the Kerr nonlinearity-induced parametric oscillations and interactions for broadband optical frequency conversion.
After the first demonstration of ultralow threshold optical parametric oscillators (OPO) in a high-Q toroid cavity~\cite{kippenberg2004KerrNonlinearityOpticalParametric}, there has been an ongoing effort to improve key parameters such as output power, conversion efficiency, and wavelength tunability~\cite{ji2017UltralowlossOnchipResonators,lu2019MilliwattthresholdVisibleTelecom,lu2020OnchipOpticalParametric,okawachi2020DemonstrationChipbasedCoupled,perez2023HighperformanceKerrMicroresonator,stone2022EfficientChipbasedOptical}.
Such compact and versatile devices hold immense potential for a wide range of applications~\cite{sun2023ApplicationsOpticalMicrocombs}, including optical communications~\cite{marin2017microresonator}, and photon pair generation~\cite{brydges2023integrated}.
The integration of OPOs on a chip offers several advantages, including enhanced stability, reduced footprint, and compatibility with existing semiconductor manufacturing processes~\cite{ji2023EngineeredZerodispersionMicrocombsa}.
These factors make chip-integrated OPOs highly attractive for practical applications where space constraints and cost-effectiveness are crucial considerations.
One of the key properties of the integrated OPO is the tunability of signal and idler wavelengths~\cite{lin2008ProposalHighlyTunable,sayson2017WidelyTunableOptical,sayson2019OctavespanningTunableParametric}.
However, an outstanding challenge remains: the realization of signal/idler frequency switching without relying on a broadly tunable (often exceeding units of THz) and expensive external pump laser.

Over the past decade silicon nitride, first considered as a capping layer for transistors, has envolved into the leading platform for low loss nonlinear integrated photonics including massively parallel ~\cite{riemensberger2020massively}, dual-comb ~\cite{Lukashchuk2021Dual} and chaotic LiDARs~\cite{Lukashchuk2021Chaotic}, as well as optical frequency synthesis \cite{Spencer2018-oc} and optical clocks \cite{Papp:14}.
Coupled resonator systems are a prominent example of breakthrough enabled by accessing the next level of complexity.
Such systems have demonstrated the presence of emergent phenomena~\cite{tikan2021EmergentNonlinearPhenomenaa,tikan2022protected}, operation on both sides of an exceptional point~\cite{komagata2021dissipative}, and increased the efficiency of microcomb generation~\cite{helgason2022power}. 

\begin{figure*}
    \centering
    \includegraphics[width=0.99\linewidth]{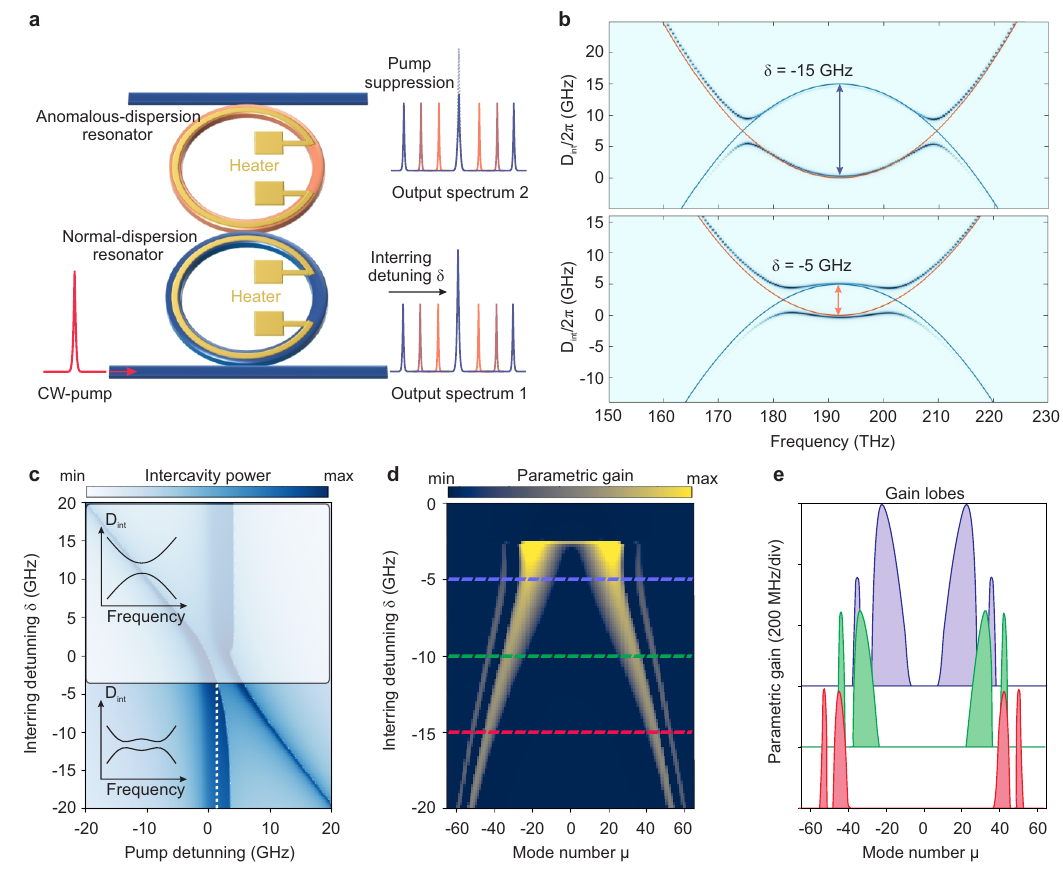}
   \caption{\textbf{Parametric gain in the hybrid-dispersion photonic dimer}. (a) Schematics of two coupled resonators with opposite signs of D$_2$. The external laser excites the system through the normal dispersion resonator. The output signal is measured from both add and drop ports. The signal collected from the drop port demonstrates the pump filtering effect. (b) Integrated dispersion hybridization for different inter-ring detunings (voltages applied to one of the heaters). (c) Intercavity power in the normal-dispersion resonator as a function of pump and inter-ring detuning.  (d,e) Gain lobes tunability for different inter-ring detunings and fixed pump detuning (dashed line in subplot c) in the hybrid photonic dimer.}
    \label{fig:pics1}
\end{figure*}

In this letter, we theoretically and experimentally investigate optical parametric oscillation in an alternating-dispersion photonic dimer, i.e. a system of two coupled optical microring~resonators~\cite{komagata2021DissipativeKerrSolitonsa,tikan2021EmergentNonlinearPhenomenaa,tikan2022ProtectedGenerationDissipativeb,roy2021NondissipativeNonHermitianDynamics,Zhang2019} with opposite signs of group velocity dispersion (GVD) (see Fig.~\ref{fig:pics1}a).
Optical coupling results in the interaction of the two fundamental mode families and the hybridization of their dispersion profiles. 
Active control over the hybridization regime in our scheme is carried out with heaters~\cite{helgason2021dissipative}.
By applying the voltage to the integrated heating element, we can dynamically adjust the curvature of the hybridized dispersion.
This causes a change in the spectral position of the signal and idler. 
Our approach enables the tunable OPO with a nearly fixed pump laser operating wavelength (signal/idler to pump tuning range 50 times exceeding previous schemes~\cite{sayson2017WidelyTunableOptical,sayson2019OctavespanningTunableParametric}), overcoming the limitations of the previous techniques.

\section{Results}
The spectral position of the fundamental-TE modes in the resonator with respect to the pumped mode $\omega_0$ can be written as:
\begin{equation}
    \omega_{\mu}=\omega_0+D_1\cdot\mu + \frac{1}{2}D_2\cdot\mu^2+...
\end{equation}
Here $\mu$ denotes the integer azimuthal mode number with respect to the pump mode $\mu = 0$. 
$D_1/2\pi$ is the free spectral range; the remaining terms are called the integrated dispersion $D_{\mathrm{int}}$. 
Depending on the sign of the dispersion parameter $D_2$, directly related to GVD~\cite{herr2014temporal}, one distinguishes two cases: positive values of $D_2$ correspond to \textit{anomalous} dispersion, and negative values correspond to \textit{normal} dispersion (see orange and blue lines in Fig.~\ref{fig:pics1}b, respectively).
If the optical inter-ring coupling strongly exceeds the loaded cavity linewidth, the optical modes hybridize and are split according to the coupling strength and relative detuning by the rate of:
\begin{equation}
    \Delta = \sqrt{4J^2 + (|D_{\mathrm{int}}^{(1)} - D_{\mathrm{int}}^{(2)}| - \delta)^2},
\end{equation}
where $J$ is the coupling rate between microresonators and $\delta$ is the relarive inter-ring detuning.
\begin{figure*}\centering
    \includegraphics[width=1\linewidth]{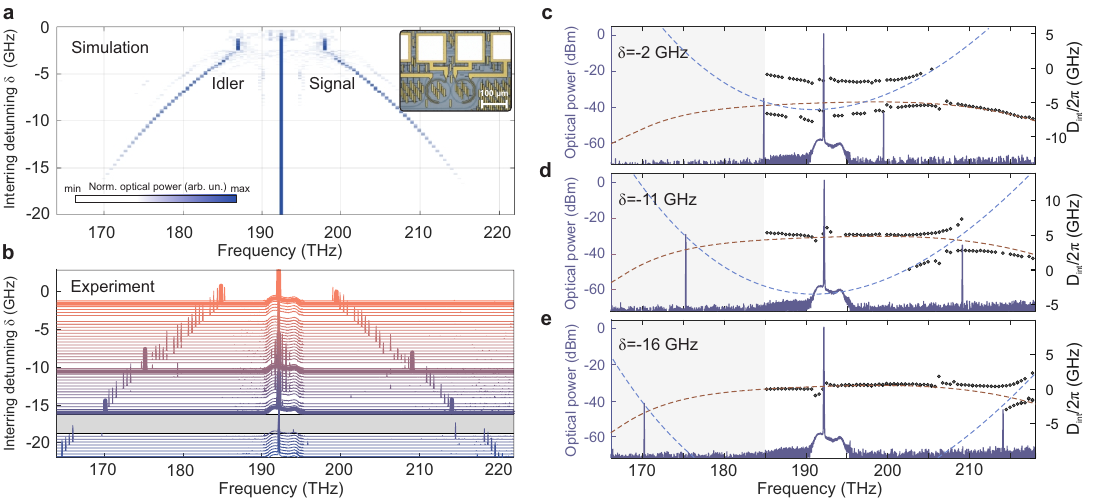}
   \caption{\textbf{Voltage-controlled OPO}. (a and b) LLE-based numerical and experimental OPO tuning by thermal tuning of the normal-dispersion resonator. Inset in (a) is a photo of the experimental device.  Thick lines in (b) are optical spectra with interring detunings $\delta$ equal to -16 GHz, -11 GHz, and -2 GHz, respectively. The gap corresponds to the region forbidden by mode crossings. (c-e) OPO spectra combined with integrated dispersion measurements for the same interring detuning.}
    \label{fig:pics2}
\end{figure*}
In general, both the coupling $J(\mu)$ and the relative detuning depend on the frequency because of the evanescent optical field decay and the difference in optical dispersion.
This gives rise to a complex dispersion landscape for the hybridized modes with many more control parameters than in the case of the single resonator.  
In our scheme, such a parameter is the inter-ring detuning $\delta$, which we control with heating elements.

An example of hybridized dispersion at different detunings is shown in Fig.~\ref{fig:pics1}b.
Here optical modes of the resonators are split most strongly in the vicinity of the crossings of the initial integrated dispersions. 
Qualitatively, it results in an additional phase-matching condition for the four-wave mixing processes that we exploit for the OPO generation.
Applying the voltage to the heaters, we shift the relative positions of the two parabolas, thereby changing the positions of the modulation instability gain lobes (see Fig.\ref{fig:pics1}d).
We analyze this effect quantitatively by examination of the Jacobian matrix eigenvalues of the two coupled Lugiato-Lefever equations (LLEs) (see Supporting Information I).  
We operate the device in the regime when the pumped mode $\mu = 0$ is strongly localized in the normal-dispersion resonator where $\Delta(0) >> \kappa$.
The simulated intracavity power for different interring detunings $\Delta$ and pump frequencies $\omega_p$ is shown in  Fig.~\ref{fig:pics1}c. 
Fixing $\omega_p$ (white dashed line in Fig.~\ref{fig:pics1}c), we calculate the position of the parametric gain lobes as presented in Fig.~\ref{fig:pics1}d,e. 
We observe that with varying inter-ring detuning $\delta$, one achieves tunability of the parametric gain and hence the spectral position of the signal/idler.

We demonstrate voltage-tunable OPO using a \textit{Si$_3$N$_4$} dimer (LGT 1 wafer, A2 chip) with FSR = $458$~GHz and $D^\mathrm{a}_2/2\pi=12.6$~MHz and $D^\mathrm{n}_2/2\pi = -1.2$~MHz for anomalous and normal resonators (see the resonator design procedure in Supplementary Information III).
First, we create a numerical model for signal/idler generation using two coupled LLEs describing the optical interaction in the two opposite-dispersion rings, using experimentally measured values of the FSR and dispersion. In our numerical experiment, we pump the normal dispersion resonator with fixed pump detuning and power while changing the detuning of the anomalous resonator. The generated signal and idler spectral position, shown in Fig.~\ref{fig:pics2}a, qualitatively matches results from Fig.~\ref{fig:pics1}d and demonstrate OPO signal tunability of 20~THz with fixed pump laser frequency. 

Experimental measurements are performed on hybrid photonic dimers fabricated by LIGENTEC SA (see inset in Fig.~\ref{fig:pics2}c) using AN800 technology.
Microheaters are designed such as not to cover the resonators coupling regions and minimize thermal crosstalk. 
All experimental data presented in Fig.~\ref{fig:pics2}b are obtained using a single experimental device (see the experimental setup in Supplementary Information IV).
We observe a broad range of operating wavelengths and precise control over the OPO's signal/idler position.
Gaps in Fig.~\ref{fig:pics2}(b) depict the inter-ring detuning range where sidebands are generated in the enhanced mode crossings~\cite{tikan2022protected}. 
In contrast to numerical simulations, the experimental demonstration of the tunable OPO required an adjustment of the pump frequency due to the fixed-by-design initial relative interring detunings of the resonators. 
Therefore, to tune the signal and idler, we heated the normal dispersion resonator.
This resulted in the need to tune the pump frequency in the $20$ GHz range.
Nevertheless, the achieved signal/idler operating range is $\approx 20$~THz.
Thus, the laser tunability requirements are orders of magnitude lower than that required by previous schemes. 

In order to directly show the correspondence between signal/idler frequency tuning and dispersion hybridization regime, we performed dispersion characterization using the frequency comb-calibrated diode laser spectrometer \cite{del2009frequency}. 
The measurements are carried out with all the same inter-ring detuning values as shown in Fig.~\ref{fig:pics2}b.
The results for selected interring detunings are shown as dots in Fig.~\ref{fig:pics2}c-e, and juxtaposed with the corresponding optical spectra (thick lines in Fig.~\ref{fig:pics2}b).
To highlight the hybridization effect on the dispersion profile, we added the experimentally fitted dispersion in the uncoupled regime ($\delta>$100 GHz) depicted with dashed lines (the fitting procedure is described in SI IV).
Thus, we clearly see that signal/idler lines are formed in the vicinity of the strongest mode interaction region and follow it when we change the inter-ring detuning.

\begin{figure}
    \centering
    \includegraphics[width=\linewidth]{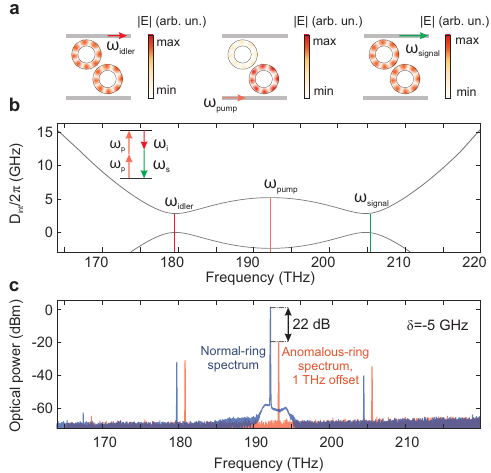}
   \caption{\textbf{OPO pump suppression in the drop-port output.} (a) Layout for pump suppression effect. At the pump frequency, the optical power is localized in the normal-dispersion ring. In the vicinity of the strongest mode interaction, power is distributed equally. (b) Schematic representation of hybridized dispersion with pump, signal, and idler frequencies separation. (c) The experimental optical spectrum for both rings at $\delta$ = -5 GHz. Blue and red colors indicate the add/drop port measurements of the optical spectrum, respectively. }
    \label{fig:pics3}
\end{figure}

One of the key applications that can be influenced by the voltage-tunable OPO is quantum communication \cite{brydges2023integrated}. In this regard, the presence of a strong pump often represents an issue for sensitive quantum detectors. Our scheme provides the possibility of (potentially strong) pump filtering. Two factors contributing to this process are pump mode hybridization and drop port employment.
While the strong mode hybridization modifies the curvature of the dispersion at the distinct mode numbers and $\delta>>J$ (see Fig.~\ref{fig:pics3}b in the vicinity of 180~THz and 205~THz), the field at the weakly hybridized pump mode is primarily located in the normal dispersion resonator (see central sketch Fig.~\ref{fig:pics3}a).
Conversely, in the vicinity of the points of strongest interaction, the optical power leaks to the auxiliary resonator. This allows the detection of the signal/idler lines with the suppressed pump on the second ring drop port. We calculate the dependence of pump suppression on the interring detuning as shown in Supplementary Information II.
However, in the fabricated device, the location of the drop port led to a cross-talk effect at the output.
The experimental result in Fig.~\ref{fig:pics3}c depicts the constant pump suppression of approximately 22~dB and partial enhancement of the sideband power.

Finally, we note that there are immediate improvements that can be envisaged in the system to further boost its performance. First, integrated actuators will completely eliminate the thermal cross-talk between the resonators as well as the requirement for pump laser tunability. Second, further optimization of the dispersion profile could extend the operation bandwidth up the octave-spanning operation regimes and result in further pump suppression at the drop port. Third, further optimization of the bus-to-ring coupling will provide enhanced efficiency of the OPO. Thus, we believe that our proof-of-concept device can evolve into a reliable tool for modern photonic applications.


\section{Conclusion}
In conclusion, the development of a broadband voltage-tunable OPO integrated on a chip represents a significant milestone in the field of integrated photonics.
Our scheme, employing two coupled resonators with opposite dispersion, enables exceptional tunability without the need for a broadly tunable and expensive laser source. 
This scheme paves the way for the realization of compact, cost-effective, and highly tunable OPOs that can be seamlessly integrated into various photonic systems.
Furthermore, the ability to achieve broadband voltage tunability without the need for a broadly tunable laser represents a significant leap forward in the field of integrated photonics. 
This advancement not only eliminates the reliance on costly external laser sources, but also enhances the device's overall efficiency and performance.
\section{Acknowledgements}
This material is based upon work supported by the Air
Force Office of Scientific Research under award number
FA9550-19-1-0250. The authors thank Tiffany Brydges for the fruitful discussion on the pump filtering section of the manuscript.

\bibliography{Main}

\end{document}


\preprint{APS/123-QED}
\title{Supplementary information: Voltage-tunable OPO with an alternating dispersion dimer integrated on chip}


\author{Dmitry Pidgaiko$^{1,2}$}
\author{Aleksandr Tusnin$^{1,2}$}
\author{Johann Riemensberger$^{1,2}$}
\author{Anton Stroganov$^{3}$}
\author{Alexey Tikan$^{1,2}$}%
 \email{alexey.tikan@epfl.ch}
\author{Tobias J. Kippenberg$^{1,2}$}%
 \email{tobias.kippenberg@epfl.ch}
\affiliation{%
 $^1$Institute of Physics, Swiss Federal Institute of Technology Lausanne (EPFL), Lausanne, Switzerland\\
 $^2$Center for Quantum Science and Engineering, EPFL, Lausanne, Switzerland
}%

\date{\today}

\maketitle
\onecolumngrid
\section{Description of the parametric gain}

The coupled Lugiato-Lefever equation can be written in the following way:
\begin{align}
\frac{d A}{d t}&=-\big(\kappa_A/2 + i (\delta \omega + \Delta_A)\big)A + i\frac{D_2^A}{2}\frac{\partial^2 }{\partial \varphi^2}A + i J B + ig_0^A |A|^2 A + \sqrt{\kappa_\mathrm{ex}^A}S_\mathrm{in}^A, \\
\frac{d B}{d t}&=-\big(\kappa_B/2 + i (\delta \omega + \Delta_B)\big)B + i\frac{D_2^B}{2}\frac{\partial^2 }{\partial \varphi^2}B+ i J A + ig_0^B |B|^2 B + \sqrt{\kappa_\mathrm{ex}^B}S_\mathrm{in}^B,
\label{eq:cLLEs}
\end{align}
where $A$ ($B$) describes the optical field envelope in the resonator with anomalous (normal) dispersion; $\kappa_{A(B)} = \kappa_0 \kappa_\mathrm{ex}^{A(B)}$ is the total linewidth composed of the internal linewidth $\kappa_0$ and the coupling to the bus waveguide; $\delta\omega$ is the laser-cavity detuning; $S_\mathrm{in}^{A(B)}=\sqrt{P_\mathrm{in}^{A(B)}/\hbar\omega}$ is the pump term with $P_\mathrm{in}^{A(B)}$ as an input power; $D_2^{A(B)}$ is the group-velocity dispersion; $J$ is the coupling strength between two resonators; $g_0^{A(B)}$ is the single-photon Kerr frequency shift. 

To analyze the modulation instability (MI) gain lobes, we use the following Ansatz
\begin{align}
    A &= A_0 + a_\mu(t)e^{i\mu\varphi} + a^*_{-\mu}(t)e^{-i\mu\varphi},\\
    B &= B_0 + b_\mu(t)e^{i\mu\varphi} + b^*_{-\mu}(t)e^{-i\mu\varphi},
\end{align}
where the CW solution $A_0$ are $B_0$ supposed to be constant in time and obey the equation
\begin{align}
   0&=-\big(\kappa_A/2 + i (\delta \omega + \Delta_A)\big)A_0 + i J B_0 + ig_0^A |A_0|^2 A_0 + \sqrt{\kappa_\mathrm{ex}^A}S_\mathrm{in}^A \label{SIeq:stable_A}\\
   0&=-\big(\kappa_B/2 + i (\delta \omega + \Delta_B)\big)B_0 + i J A_0 + ig_0^B |B_0|^2 B_0 + \sqrt{\kappa_\mathrm{ex}^B}S_\mathrm{in}^B.\label{SIeq:stable_B}
\end{align}
The amplitudes $a_\mu$ and $b_\mu$ are assumed to be small enough so their dynamics in time can be linearized in the vicinity of the stable solutions $A_0$ and $B_0$. The resulting linearized system of equations takes form
\begin{align}
\dot{a}_\mu &= -\big(\kappa_A/2 + i (\delta \omega + \Delta_A)\big)a_\mu - i \frac{D_2^A}{2}\mu^2 a_\mu + i J b_\mu +i g_0^A(2|A_0|^2a_\mu+A_0^{2^*}a_{-\mu}^*)\\
\dot{a}^*_{-\mu} &=-\big(\kappa_A/2 - i (\delta \omega + \Delta_A)\big)a^*_{-\mu} + i \frac{D_2^A}{2}\mu^2 a^*_{-\mu} - i J b^*_{-\mu} -i g_0^A(2|A_0|^2a^*_{-\mu}+A_0^{2}a_{\mu})\\
\dot{b}_\mu &= -\big(\kappa_B/2 + i (\delta \omega + \Delta_B)\big)b_\mu - i \frac{D_2^B}{2}\mu^2 b_\mu + i J a_\mu +i g_0^B(2|B_0|^2b_\mu+B_0^{2^*}b_{-\mu}^*)\\
\dot{b}^*_{-\mu} &=-\big(\kappa_B/2 - i (\delta \omega + \Delta_B)\big)b^*_{-\mu} + i \frac{D_2^B}{2}\mu^2 b^*_{-\mu} - i J a^*_{-\mu} -i g_0^B(2|B_0|^2b^*_{-\mu}+B_0^{2}b_{\mu}).
\end{align}
This system of equations can be rewritten in matrix form for the unknown vector $X = (a_\mu, a_{-\mu}^*,b_\mu, b_{-\mu}^*)^T$ in the form
\begin{equation}
    \dot{X} = \mathbb{M}X.
\end{equation}
The parametric gain rate for the OPO sidebands can be then inferred from the analysis of the eigenvalues $\lambda_j$ (j=$1,2,3,4$) of the matrix $M$. To compute this matrix, we solve numerically equations~(\ref{SIeq:stable_A},\ref{SIeq:stable_B}) imitating experimental tuning of the pump laser from the blue to red-detuned zones of the resonances.

\begin{figure}
    \centering
    \includegraphics[width=\textwidth]{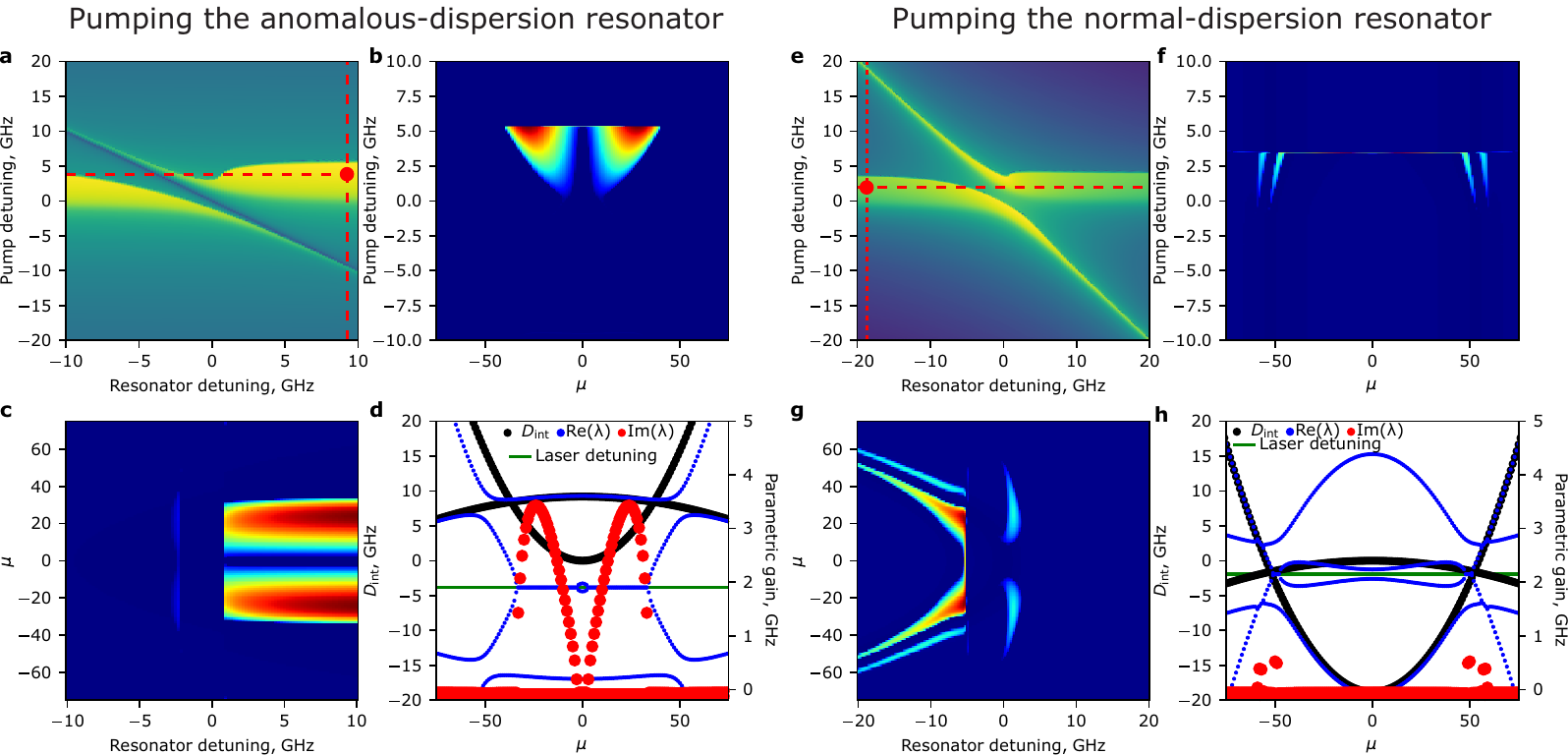}
    \caption{\textbf{Modulation instability gain lobes.} a) Field response in one of the resonators for different pump and resonator detunings. b) MI gain lobes as a function of pump detuning along vertical line in a). c) MI gain lobes along the horizontal line in a). d) Integrated dispersion and eigenvalues $\lambda_j$   structure for the point in panel a). (e-f) The same, but for the excitation of the normal dispersion resonator. }
    \label{SIfig:gain_lobs}
\end{figure}

In Fig.~\ref{SIfig:gain_lobs} we present the computed MI gain lobes and the structure of the eigenvalues $\lambda_j$. To compute it, we used $\kappa_0/2\pi = 60$~MHz, $D_2^A/2\pi = 12.6$~MHz, $D_2^B/2\pi = -1.1$~MHz, $\kappa_\mathrm{ex}^A = \kappa_0$, $\kappa_\mathrm{ex}^B = 2\kappa_0$, $P_\mathrm{in}^{A(B)} = 0.3$~W. In Fig.~\ref{SIfig:gain_lobs}(a-d) we consider the excitation of the anomalous dispersion resonator. We observe, that for the constant resonator detuning $\Delta_B$ ($\Delta_A = 0$), the MI gain lobes correspond to the conventional single-resonator case (see Fig.~\ref{SIfig:gain_lobs}b). As shown in Fig.~\ref{SIfig:gain_lobs}c, the gain lobes almost do not change their position as they are dominated by the conventional Turing instability in the anomalous dispersion (see Fig.~\ref{SIfig:gain_lobs}d). However, when we pump the normal dispersion ring (see Fig.~\ref{SIfig:gain_lobs}(e-f)) and tune $\Delta_A$ ($\Delta_B=0$), we observe the desired effect. Since we pump the region with normal dispersion, Turing instability does not manifest, and the sidebands experience positive parametric gain due to the new phase-matching conditions. For the fixed pump detuning, the gain lobes change their position as shown in Fig.~\ref{SIfig:gain_lobs}g. The eigenvalues in Fig.~\ref{SIfig:gain_lobs}h show, that the new phase matching condition occurs due to the crossing of the anomalous and normal dispersion parabolas. Thus, one can estimate the position of the modes with positive parametric gain via simple formula
\begin{equation}
    \mu = \pm \sqrt{\frac{2 \Delta_A}{D_2^A-D_2^B}}.
\end{equation}
The latter estimation works in the limit $\Delta \gg J$. To achieve bigger tunability, one needs to reduce the group velocity dispersion of both resonators.

\section{Pump suppression}

Let us estimate the lower limit of the pump suppression.
For simplicity, we will introduce a normalized version of the LLE considering a CW-type solution and neglecting the nonlinear terms:
\begin{align}
\frac{\mathrm{d}}{\mathrm{d}t} \psi_a &= -[1 + i (\xi_0+\tfrac{1}{2}\delta_n )]\psi_a  + i  J_n \psi_b + f \notag\\
\frac{\mathrm{d}}{\mathrm{d}t} \psi_b &= -[\kappa_n + i (\xi_0-\tfrac{1}{2}\delta_n)]\psi_b  + i  J_n \psi_a,
\label{eq:cLLEs_norm}
\end{align}
here the subscript index $n$ stands for normalized by $\kappa_A/2$, $\kappa_n = \kappa_B/\kappa_A$, $J_n = 2J/\kappa_A$.

Equivalently, we can re-write Eq.~\ref{eq:cLLEs_norm} in the matrix form considering the stationary case in the symmetrically coupled dimer:
\begin{equation}
 \begin{array}{c}
\begin{pmatrix}0\\0\end{pmatrix}
 \end{array}
 = \mathcal{M}
\begin{pmatrix}\psi_a\\\psi_b\end{pmatrix}
+
\begin{pmatrix}f\\0\end{pmatrix},
\end{equation}
where 

\begin{equation}
  \mathcal{M} =  \left(
\begin{array}{cc}
 -1-i \left(\xi_0 +\frac{\delta_n}{2}\right) & i J_n \\
 i J_n & -1-i \left(\xi_0 -\frac{\delta_n}{2}\right)
\end{array}
\right).
\end{equation}

Thus, the solution of this system is given by:
\begin{equation}
 \begin{array}{c}
\begin{pmatrix}\Tilde{\psi}_a\\ \Tilde{\psi}_b\end{pmatrix}
 \end{array}
 = -\mathcal{M}^{-1}
\begin{pmatrix}f\\0\end{pmatrix}=
\begin{pmatrix} -\frac{f \left(-1-i \left(\xi_0 -\frac{\delta_n}{2}\right)\right)}{\frac{\delta_n^2}{4}+J_n^2-\xi_0 ^2+2 i \xi_0 +1}\\\frac{i f J_n}{\frac{\delta_n^2}{4}+J_n^2-\xi_0 ^2+2 i \xi_0 +1}. \end{pmatrix}
\end{equation}

We define pump suppression in the following way:
\begin{equation}
    \eta = \left|\frac{S_\mathrm{out,1}}{S_\mathrm{out,2}} \right|^2= \frac{\left| f-\psi_a\right|^2}{\left|\psi_b\right|^2}.
\end{equation}
Taking normalized field amplitude at the center of the pumped resonance, we obtain:
\begin{equation}
    \eta =  \frac{3 \delta_n \sqrt{\delta_n^2+4 J_n^2}+5 \delta_n^2+18 J_n^2+8}{2 J_n^2}.
    \label{eq:pump_suppr}
\end{equation}

\begin{figure}
    \centering
    \includegraphics[width=0.5\linewidth]{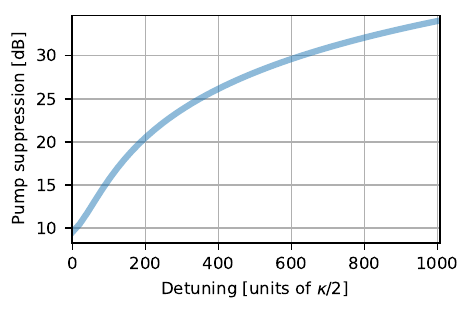}
   \caption{\textbf{Pump suppression at the drop-port output.} }
    \label{fig:si2}
\end{figure}

The lower limit of the pump suppression given by Eq.~\ref{eq:pump_suppr} is depicted in Fig.\ref{fig:si2}. The pump detuning is chosen to be at the position of the resonance. One can see that in the case when $\delta_n \gg J_n$ Eq.~\ref{eq:pump_suppr} can be simplified as $\eta \approx 4\delta_n^2/J_n^2$. It reaches the value of 34 dB in the case of large detuning which corresponds to the largest signal-to-pump frequency offset presented in Fig.~2b of the main article.

\section{Resonators design}
\begin{figure}[b]
    \centering
    \includegraphics[width=\linewidth]{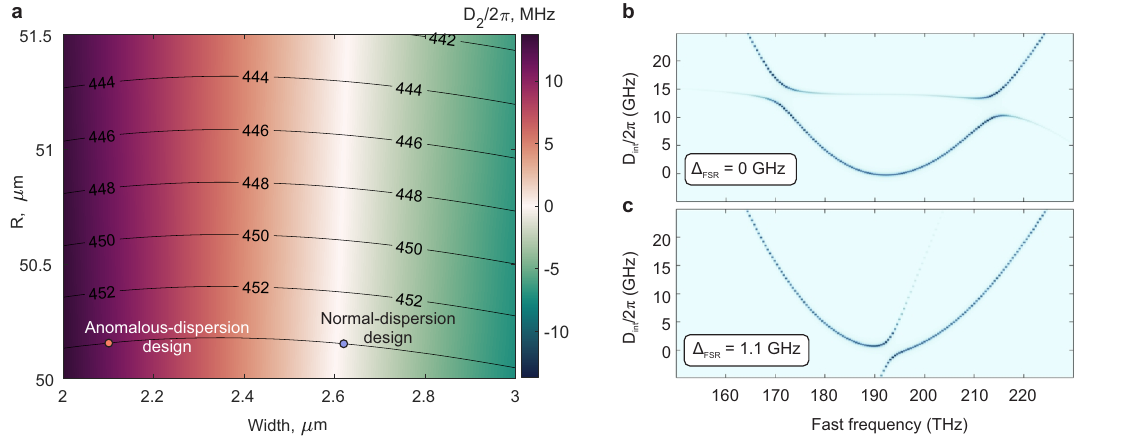}
   \caption{\textbf{a} Normalised dispersion parameters map for different resonator widths and radii, and fixed height.  Black lines are equal-FSR lines (labeled in GHz). Dots highlights the final design.\textbf{b} Hybridised dispersion for the chosen resonator parameters. The top figure shows the case with equal FSRs. In the lower one, the FSR difference of 1.1 GHz is artificially introduced.}
    \label{fig:Design}
\end{figure}

The search for the geometric parameters of the resonators with different dispersion type was carried out with COMSOL Multiphysics.
In the simulation, the radius and width of the micro-ring were varied, height was fixed.
The simulated parameter is the effective refractive index for the fundamental TE mode, from which we calculated FSR and $D_2/2\pi$. 
The map with $D_2/2\pi$ is shown in Fig.~\ref{fig:Design}a.
The black lines indicate equal FSR in GHz.
The choice of the final design was based on two requirements. First, the rings must have a different $D_2$ sign.
Second, they must have equal FSR. 
The importance of FSR equality is explained in Figs.~\ref{fig:Design}b and c.
Hybridized dispersion with perfectly matched FSR is plotted in Fig.~\ref{fig:Design}b, the result in Fig.~\ref{fig:Design}b is calculated for dispersions with 1.1 GHz FSR difference.
Even such a small difference (less than 1 percent of FSR) leads to a deviation of the hybridization mode from the required.

\section{Experimatal setup and uncoupled dispersion}
\begin{figure}
    \centering
    \includegraphics[width=\linewidth]{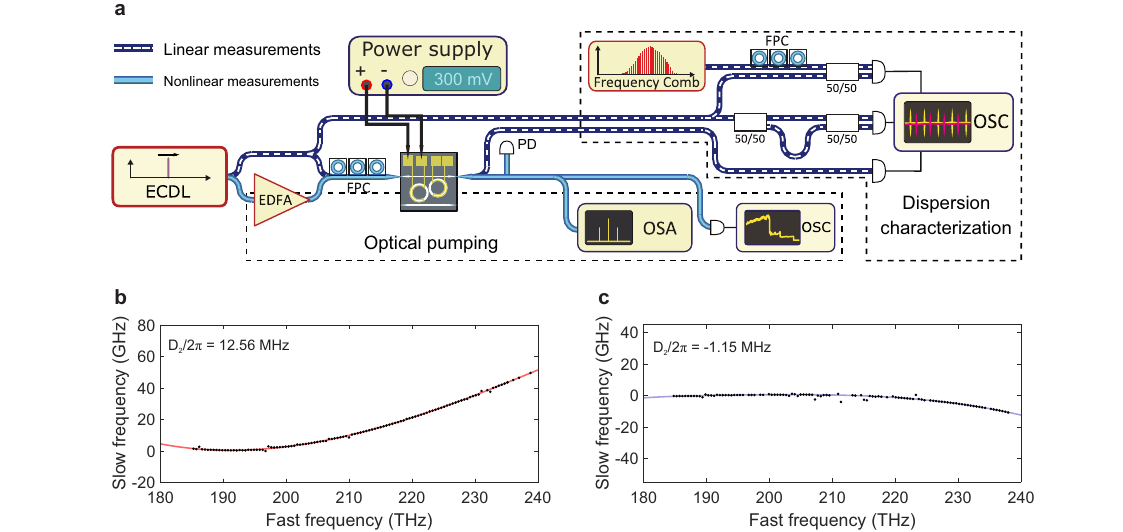}
   \caption{\textbf{a} Experimental setup. \textbf{b,c} Fitted integrated dispersion (solid lines) for dispersion characterization data (dots) for anomalous- and normal-dispersion resonators, respectively.}
    \label{fig:Setup}
\end{figure}

Our experimental setup operates in two regimes. 
The first (linear measurements) is dispersion characterisation~\cite{del2009frequency} where one can measure FSR and integrated dispersion.
The second regime is intended for nonlinear optical pumping and measurement of optical spectra of generated light.
In this channel, the pump with a tunable wavelength from the external-cavity diode laser is enhanced by an erbium-doped fiber amplifier and then coupled to a dimer. 
The generated signal is then analyzed with an optical spectrum analyzer.
The setup includes a photodiode connected to an oscilloscope to measure transmission traces. 
In both regimes, the microrings' heater elements are supplied with DC power source, which is necessary to tune the interring detuning.
With zero power on the heaters, the detuning is so great that one can consider the rings optically uncoupled. 
The integrated dispersion measurements for normal and anomalous rings at zero power are shown in Fig.~\ref{fig:Setup}b and c. We carried out all the experiments with LGT1 wafer with A2 chip.  According to a fit, $D_2/2\pi$ values are 12.56 MHz and -1.15 MHz, respectively. The measured FSR deviation is 0.1 GHz. The fitting curves are used for the analysis of hybridization regimes in Fig. 2c-e in the main paper. 

\section{Detunning calibration}
\begin{figure}[b]
    \centering
    \includegraphics[width=0.85\linewidth]{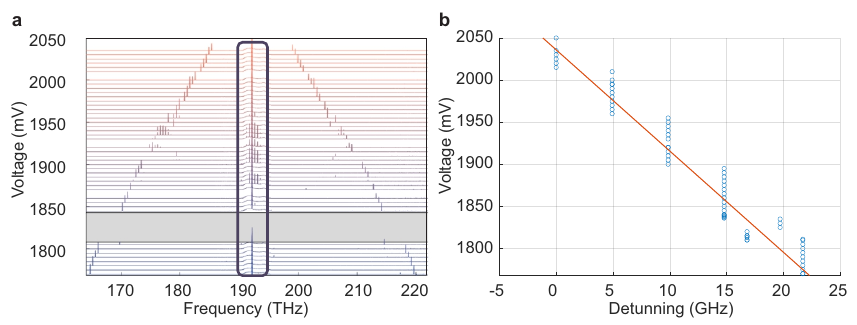}
   \caption{\textbf{a} Experimental spectra for different voltages. \textbf{b} }
    \label{fig:Scale}
\end{figure}

Experimental data on the detuning to heters voltage relation are contained directly on the main experimental result shown in Figure 2b of the main paper. Each spectrum corresponds to a different heater voltage in this experiment, so the initial result has a voltage on the y-axis (see Fig.~\ref{fig:Scale}a). Since we were pumping the same ring as we were heating during the experiment, we needed to adjust the frequency of the pump laser each time while tunning voltage. Thus, by tracking the shift of the pump laser (the center lines of the spectra in Fig.~\ref{fig:Scale}a), we can extract the graduation scale from voltage to absolute frequency value. In order to translate this data into relative interring detuning, we need an initial condition that we have chosen a voltage value that provides zero relative detuning between dispersions (2050 mV). The result is shown in Fig.~\ref{fig:Scale}b, for which we took a linear approximation and used it to recalculate each voltage from the Fig.~\ref{fig:Scale} and got main result shown in Fig.2b of the main article. 
\bibliography{apssamp}